\documentclass[aps,prc,preprint,showpacs]{revtex4-1}
\usepackage{bm}
\usepackage{color}
\usepackage{xcolor,graphicx}

\usepackage{dcolumn}
\usepackage{amsmath}
\usepackage{amsfonts}
\usepackage{amssymb}
\newcommand{\be}{\begin{equation}}
\newcommand{\ee}{\end{equation}}
\newcommand{\bea}{\begin{eqnarray}}
\newcommand{\eea}{\end{eqnarray}}
\begin{document}
\title{Optimal Intrinsic Frame of Reference for Deformed Nuclei}
\author{N. A. Lyutorovich}
\email{lyutor@nuclpc1.phys.spbu.ru}
\affiliation{Department of Nuclear Physics, Physical Faculty, 
St. Petersburg State University, 198504, St. Petersburg, Russia}
\date{\today}

\begin{abstract}

Nowaday, in study of effective interactions, more attention is devoted 
to single-particle properties of near-magic nuclei 
and bulk properties of deformed ones but quasiparticle states
of the latter are rarely used so far because of theoretical difficulties.
In particular, the angular momentum projection remains too time-consuming
for such calculations and the methods, which are based on the transformation
to an intrinsic frame have some unsolved problems such, e. g.,
as quantum fluctuations of rotational recoil in the description
of quasiparticle-rotation coupling. To remove a part
of these difficulties, the method of the optimal intrinsic 
frame-of-reference is developed. After applying
the Mikhajlov transformation to obtain the
nuclear Hamiltonian in the intrinsic frame,
approximate constraints on nucleon's variables are substantiated, and
the quasiparticle structure of the nucleus orientation angle operators
is investigated. That gives possibility to use the variational principle
to derive equations for matrix elements of these operators.
An approximation similar to the cranking model (CM), but
with the quantum rotational recoil, is formulated, which may be considered
as a generalization of the usual self-consistent CM.
Simplified model calculations for rotational bands in
$^{163}$Er show that taking into account the recoil operator 
considerably improves the agreement with experimental data.
\end{abstract}


\maketitle

%
%
\section{Introduction}
%
When studying nuclei far away from the stability valey, a need arises
to improve the methods of the microscopic description
of deformed nuclei since, on the one hand, many nuclei in
the region are deformed, and on the other hand,
the semi-microscopic methods using phenomenological
single-particle potentials and residual interaction are often
inapplicable here because of lack of data for determining
the model parameters. The microscopic description of traditional regions
of deformed nuclei, near the stability valey, has a long history
(see, e.~g., books \cite{BMII} - \cite{RS}, 
reviews \cite{Ring_1985}--\cite{Vretenar_2005}, and references therein)
with the latest time witnessing great progress by virtue of improvements
in computer methods. However, even for these regions, microscopic methods
are applicable with considerable restrictions;
in addition, a amount of self-consistent
microscopic calculations for heavy nuclei is small so far,
especially as compared with the huge amount of existing experimental material.
The development of microscopic
methods for such nuclei is of interest for many reasons, and
in particular, because this opens up new possibilities
for studying the nuclear effective interactions.

At present in study of effective interactions and even
in ab-initio calculations, the apparent
shift of attention from ground-state bulk to single-particle properties
of magic and near-magic nuclei occurs (see, e. g.,
\cite{Zalewski_2008}--\cite{Barbieri_2009}). At the same time, when
determining interaction parameters from properties of ground states,
just the deformed nuclei rather than magic ones are
increasingly used \cite{Niksic_2008,Baldo_2010},
but the quasiparticle states of the deformed nuclei 
are rarely used so far because of theoretical difficulties.

In the deformed nuclei, particle-vibration coupling is
much weaker, in many cases, than in spherical ones as the main part of this 
coupling is included in the mean deformed field
\cite{BMII} (p. 242). Therefore, along with vibrational states, there
are many excited states of essentially pure single-quasiparticle
nature in deformed nuclei (the probability of phonon admixtures
are less than 5\% \cite{Gareev}; an evaluation of amplitudes is given, also,
in Ref. \cite{BMII}), that makes, to some extent, 
these nuclei similar to the magic ones. 
The treatment of the superfluidity, which gives single-quasiparticle spectra
in deformed nuclei (instead of single-particle ones in the magic nuclei),
is elaborated enough well. But in addition, there exists 
the quasiparticle-rotation interaction in the deformed nuclei.
The matrix elements (ME) of this interaction depend on the
effective forces much weaker than the ME of the quasiparticle-phonon 
interaction in spherical nuclei, therefore the influence 
of the forces is more transparent.
The main difficulties in theoretical description of these nuclei are
related to the restoration of the broken rotational symmetry, that is
a serious obstacle in self-consistent microscopic calculations.
Therefore the overcoming these difficulties would have important advantages. 
Then the microscopic description of deformed nuclei and 
especially of odd ones could yield vast valuable information 
on effective interactions: degree of their universality, 
their dependence on nuclear shape and excitation.

For the microscopic description of deformed nuclei,
there are used two ways: the first method exploit wave functions 
with broken symmetry (as a leading approximation or as some building elements),
the second one works with wave functions which on every stage
have symmetries of the nucleus Hamiltonian, are used.
The ways are described in many good reviews, only some of them
are pointed out here: \cite{Ring_1985}--\cite{Vretenar_2005}.
The brief listing of methods with a description of their 
advantages and drawbacks may be found, e.g., in Ref.~\cite{Egido_LNP_2004}.
Supplementing these reviews it worth mentioning the self-consistent 
collective coordinate method and other approaches within the framework
of time-dependent Hartree-Fock (HF) method 
(see \cite{Hinohara_2008,Shimizu_2008} and references therein)
and also the interacting boson model, combined with the cranking one,
\cite{Nomura_2011}, pseudo-SU(3) model \cite{Vargas_2002}.
Some of publications, more closely related to the given investigation,
are mentions below in the given paper.

Since deformed potentials enable to allow for an essential part of
multinucleon correlations in a simple form, the approaches 
that use the symmetry violation prove to be most efficient. 
Among the methods using such approaches, the angular momentum projection (AMP) 
and the intrinsic frame-of-reference (IF)  are basic \cite{RS}.
Both AMP and IF methods bear on some common physical ideas, 
in particular, the concept of intrinsic wave functions and deformed potentials
in IF (the projection of wave functions can be interpreted as a transformation 
of the functions to IF, complimented with integration over all possible 
orientations of the frame). 
Nevertheless,  mathematical formulations of the methods differ markedly.
To date, AMP is much more developed and applied than IF. 
The development of computers has made possible self-consistent
calculations by configuration mixing of states obtained by variation
after AMP. Only some publications of such works performed in the frame work 
of both the nonrelativistic and relativistic approaches,
will be pointed out here: \cite{Bender_2008}--\cite{Ring_2011}.
Nonetheless, further development of microscopic theory
of deformed nuclei is needed, since the application
of self-consistent AMP method is so far possible only
with considerable restrictions, especially it relates to describing 
the interaction of single-quasiparticle and rotational degrees
of freedom in heavy nuclei.

The second of the mentioned approaches using the symmetry violation,
the IF method, finds practically no application in
self-consistent microscopic calculations. Moreover, despite
numerous studies (see, e. g., \cite{RS}, \cite{Bes86}--\cite{LM98} and
references therein) and the apparent, at first sight, physical
obviousness of the method, the description of a nucleus in IF
encounters still with serious unsolved problems which attenuate
the activity in the direction. As a result, the basic
methods for heavy odd nuclei are still semi-classic
or semi-microscopic ones. These are the cranking model (CM) and
its more general version, tilted- axis CM (see reviews
\cite{Frauendorf_2001,Bender_RMP_2003} and latest papers
\cite{Peng_2008}--\cite{Schunck_2011}) and also a many-particle
plus rotor model (e. g., \cite{Carlsson_2007,Qi_2009}).

In the present paper, an attempt is made to overcome 
some of the difficulties of the IF method. This can be useful 
for the formulation of an efficient approach to describe 
the quasiparticle-rotation interaction in nuclei
(one of advantages of IF method is that it, unlike the AMP, requires no calculation 
of overlap integrals), for better understanding of the relationship
between the AMP and IF, and for the derivating quantum corrections
to widely used models, first of all, to the CM.
The neccessity for such corrections calls for a special explanation.

In the low approximation the AMP and IF methods may be reduced
to the self-consistent CM (SCM). Within the framework of the first method,
the SCM equations are derived on the basis of the Kamlah 
expansion \cite{Kamlah} for the expectation value of the nuclear Hamiltonian
over the projected wave functions.
A series arising here descends in the degree of coherence
of its terms \cite{RS}.
The second method gives as a series not for the expectation value but the
nuclear Hamiltonian, transformed to variables of rotating frame
of reference \cite{LM98}.
Both methods derive the SCM equations with taking account of only 
two first terms of the corresponding series. One of the terms describes,
conventionally speaking, the intrinsic motion, since it is independent
of rotational variables, and the second one describes 
the quasiparticle-rotation interaction.
Conventionality of such terminology consists in the fact that 
the expectation value of the Hamiltonian is found by wave functions,
allowing for rotation; moreover, the effective nucleonic forces
may depend on nuclear density, which itself may depend on angular
momentum of the nucleus. The rest of the terms, ignored in the SCM,
describe the rotational recoil effect, more precisely,
its fluctuation part, since some averaged part of the effect is
accounted for in the quasiparticle-rotation interaction of the SCM
(see, e. g., \cite{Hilton}). In the other terminology, 
the rotational recoil effect, like a similar effect for translations,
corresponds to subtraction of the spurious motion in IF.

Importance of the quantum recoil effect has been discussed many times
(see, e.~g., Refs.~\cite{Frauendorf_1969,Jansen_1979,Hilton,Ring_1985});
yet, majority of works that use the SCM ignores the effect.
Corrections to the SCM, derived on the AMP basis in Ref.~\cite{Une_2001},
are applicable only to states with large values of the quantum number $K$,
i.~e., the projection of the angular momentum of the nucleus 
to the symmetry axis. An efficient means to correct the SCM on the
AMP basis is the projecting of cranked states
(see Refs.~\cite{Zdunczuk_2007,Dobaczewski_2009,Schunck_2011}
 and references therein). Here, the projection before variation,
necessary to reproduce the correct moment of inertia,
is replaced by its first approximation, SCM, and the projection
(after variation) of obtained functions gives quantum corrections
to the SCM. However, such approach can so far be implemented
only in the HF approximation without pairing correlations 
and does not take into account an influence of the quantum rotational
recoil on mean fields, which is essential in both particle-hole
and particle-particle channels.

For odd nuclei, probably, for the present there is no
publication of a calculation, allowing for the quantum effect
of rotational recoil on the basis of the SCM or a generalization of it.
At the same time, as calculations have
shown \cite{LM95}, the SCM yields some underestimation of
quasiparticle-rotation coupling as compared with the observed
one. The difference between the model used in \cite{LM95}
and the SCM is immaterial for this conclusion
(the effect of the perturbative treating of 3-quasiparticles states
in the framework of SCM was considered in Ref.~\cite{Ring_1974}).
The underestimation is quite opposite, though less by magnitude, to
the well-known overestimation of the Coriolis coupling in the
particle-rotor model \cite{BMII}. It shows up in theoretical
rotational level energies for strongly-mixed bands which, in
the SCM, exceed the experimental ones with the differences
increasing for the increasing nucleus spin $I$. Note that, 
at first sight, the application of the SCM in the
independent-quasiparticle approximation to description of
positive parity band in $^{155,159,161}$Dy leads to agreement
with experiment, which has been demonstrated
in Ref.~\cite{Ring_1974}, and was confirmed by our calculations.
However, the agreement arises due to ignoring another,
important for  these states, effect. The matter is that, in
$^{155,159,161}$Dy, the states of intrinsic motion, originating
from shell $i_{13/2}$, contain large quasiparticle-vibrational
admixtures ($\,\gtrsim \,20$\% in states $3/2^+$[651] and
$7/2^+$[633] \cite{Gareev}\,). When the admixtures are
accounted for, the ME of the quasiparticle-rotation interaction
decrease (e. g., for $^{161}$Dy, such attenuation of the
interaction was discussed in the framework of the
particle-rotor model \cite{Kvasil})
and the agreement of the SCM with experiment is broken.

Let us return to the description of goals and features of the present work.
The work develops the method of optimal IF (OIF),
based on the Mikhajlov
\footnote{In some articles, written by V. M. Mikhajlov, 
his name was translated as 'V. M. Mikhailov'.}
transformation of wave functions
and operators from the laboratory coordinate system to a rotating one
(see \cite{Mikhailov_71, Mikh93, ML94}). 
To this effect, approximate conditions of constraint on nucleonic
variables are substantiated,
the quasiparticle structure of the nucleus orientation angle operators
is studied, the equations for ME of these operators are derived, 
and an approximation is stated, 
which can be considered as a CM generalization,
containing the quantum effect of the rotational recoil.
The results obtained are intended for odd nuclei, though
most part of them can be useful for even ones, too.

The work sets no goal to carry out microscopic calculations for specific
nuclei, as this is a separate serious problem. The calculations
presented in Section~\ref{sect4} pursue another aim of no less
importance: to analyze main features of the suggested approach
and reveal the role of different effects, starting with a
simple Hamiltonian and semi-microscopic approximation, which
combines diagonalization with perturbation theory.
The example $^{163}$Er is used to show that
the accounting for rotational recoil operator within the
framework of the generalized CM improves noticeably the
agreement with experiment.

The paper is organized as follows. In Sect.~\ref{sect2}, after 
a short description of the transformation to the IF,
the constrains
between nucleonic variables in this frame are considered.
Then, the general ideas of the OIF method are introduced.
The quasiparticle structure of the nucleus orientation angle operators
is studied in Sect.~\ref{sect3}. At the end of this section,
the difference between the OIF and the approximate AMP methods is shown.
The formulation of the method in the simple model and the results
of calculations in the model are given in Sect.~\ref{sect4}.
Section~\ref{sect5} contains a summary and an outlook to future 
development of the method.
\section{Transformation to the intrinsic system and constrains
between the variables \label{sect2}}
%
General ideas of the method of a rotating frame of reference are 
set forth in Ref.~ \cite{LM98} where also references to previous
works by other authors are given. The method employs V.~Mikhajlov concept 
of the unitary transformation of wave functions and operators 
from the laboratory frame to some intrinsic one that rotates with the nucleus
(see \cite{Mikhailov_71, Mikh93, ML94} and references therein).
The principle of the method is in brief as follows.
Let $x_i$, $x'_i$ and $\Psi_I( x_i)$, $\widetilde{\Psi}_I(x'_i, \vartheta)$ 
denote, respectively, nucleon variables and wave functions of the nucleus 
in the laboratory and intrinsic systems, where $i=1, 
\ldots A$, $A$ is the number of nucleons, $I$ is the nuclear spin.
We denote orientation angles of the intrinsic with respect 
to the laboratory frame as $\vartheta_a $ with understanding
$\vartheta$ to be the entire set of $\vartheta_a $
and $x_i$ to be the set of variables $\{x_i \}$. The transformation 
$\mathcal U$ is defined with the equation: 
$\Psi_I( x_i, \zeta ) = \widetilde{\Psi}_I(x'_i, \vartheta ) \equiv 
{\mathcal U} \Psi_I(x'_i, \vartheta)$. 
The additional variables $\zeta_a\ $ are introduced
for the number of variables before the transformation to be equal 
to that after it and can be interpreted as the orientation angles 
of the rotating frame of reference with respect to the axes of
the nucleus. The redundant degrees of freedom are eliminated 
with constraints on variables in the IF.

The orientation of the nucleus with respect to the rotating frame 
of reference is described in terms of $x'_i$ by three commuting operators 
$\hat{\theta}_a(x'_1,...x'_A) \equiv \theta_a(x'_i)$,
which are dependent on momentum, spin and isospin operators of nucleons 
as well as on coordinates. The selection of operators $\theta $ is carried out
from considerations of the problem simplification and is restricted 
by the requirement that their commutation rules with
the projections of the nuclear angular momentum $J_a(x'_i)$ 
be the same as those for the orientation
angles of a rigid rotor $\vartheta_a$ with operators 
$\grave{I}_a$ of the projections of its angular
momentum to the laboratory axes:
\begin{equation} \label{commutation}
[\grave{I}_a,i\vartheta_b]=b_{ab}(\vartheta) , \qquad
[J_a,i\theta_b]=b_{ab}(\theta) , \qquad
[\theta_a,\theta_b]=0\;.
\end{equation}
If $\vartheta_a$ are Euler's angles or three angular parameters 
that fix the rotation axis and the rotation angle of a rigid body, 
the expressions for $b_{ab}(\vartheta)$ can be found 
in Ref.~\cite{Varshalovich}. Below, the Cartesian components of the vector,
defined by the rotation axis and angle, are used as 
$\vartheta_a$ ($a,b=x,y,z$); 
the corresponding expressions for $b_{ab}(\vartheta)$ are given 
in Ref.~\cite{LM98}. The transformation ${\mathcal U}$ is chosen so
that the equations hold:
\begin{equation} \label{UJU}
J_\alpha (x_i)={\mathcal U} J_\alpha (x'_i)\; {\mathcal U}^{-1} =
\grave{I}_\alpha ( \vartheta) = -ib_{\alpha \mu}
\frac{\partial}{\partial \vartheta_\mu},
\qquad D(\zeta) = {\mathcal U} D(\vartheta)\; {\mathcal U}^{-1} =
D(-\theta(x'_i)),
\end{equation}
were $D(\vartheta) \equiv D^{\lambda}_{\mu \nu}(\vartheta)$ are Wigner
$D$-functions; the summation over repeating Greek indexes is implied 
here and below. This requirement leads to the expression
${\mathcal U} = $ $\exp{(\vartheta_{\alpha} J_{\alpha})} \,
\exp{(-\theta_{\beta} \grave{I}_{\beta})}\,$
\cite{Mikhailov_71}.
Note that the use of the operator angles $\theta_a(x'_i)$ presupposes
that the nuclear wave function before $\mathcal U$-transformation,
i.~e. the function in terms of the laboratory coordinates,
has the subsidiary variables on the left from nucleonic ones. 
Then, in the simplest case, the function before 
$\mathcal the U$-transformation has the form $\Psi_I( x_i,
\zeta ^{-1}) = f(\zeta^{-1})\Psi_I( x_i )$, and after it --- 
$\widetilde{\Psi}_I(x'_i,\vartheta ) = f(\theta)\,\Phi_I(x'_i, \vartheta )$. 
To be more precise, this is valid in the case of small
$\zeta$, which is discussed below, while a more involved relation,
based on Eq.~(\ref{commutation}), should be used for arbitrary $\zeta$.

In the 3D case, the nuclear Hamiltonian in the rotating frame of reference 
was obtained by Mikhajlov as an infinite series \cite{Mikh93, ML94}:
\begin{equation}
\widetilde{H} \equiv {\mathcal U} H\; {\mathcal U}^{-1} =
\sum_k \frac{1}{k!} \,[\,[\ldots[\,[H,i\theta_{\alpha_1} ],
i\theta_{\alpha_2}],\ldots], i\theta_{\alpha_k}] \,
S_{\alpha_1}\,S_{\alpha_2} \ldots S_{\alpha_k} , \label{H_Mikhajlov} \\
\end{equation}
where 
$I_{\alpha} \equiv$ $D^1_{\alpha\beta}(\vartheta^{-1})\: \grave{I}_\beta$
is the projection of the angular momentum of a rigid body 
to the intrinsic axes, 
$S_{\alpha} = b^{-1}_{\alpha\beta}(\theta)\,(I_{\beta}-J_{\beta})$,
and $b^{-1}_{\alpha\beta}(\theta)$ is the matrix inverse with respect
to $b_{\alpha\beta}(\theta)$. The series as many other expressions
in the approach in question can be radically simplified, 
if the ME of operators $\theta$ are small enough. 
Then the transformed Hamiltonian (\ref{H_Mikhajlov}) and the commutation 
relations (\ref{commutation}) for $\theta$ take the form:
\begin{eqnarray}
&& \widetilde{H} = H + \sum_a [H,i\theta_a ] (I_a-J_a) +
\frac{1}{2!} \sum_a \,[\,[H,i\theta_a ],i\theta_a] \,(I_a-J_a)^2 +
\ldots\,, \label{simple_H} \\
&& [J_a,i\theta_b]=\delta_{ab} \qquad a,b=x,y \;. \label{simple_commut}
\end{eqnarray}
The present work is confined to consideration of axially symmetrical
nuclei, therefore, operators $\theta_z$ 
are not used in the formulas (some deviations from the axial symmetry
can be allowed for by mixing states with different projections 
of the angular momentum to the symmetry axis). 
The smallness of $\theta$ is ensured by imposing constraints on variables.
If the constraints are chosen so that $\zeta_a $ were small enough before
the $\mathcal U$-transformation, the equations for $D$-functions 
in (\ref{UJU}) can be replaced with a simpler equation 
$\zeta_a = -\theta_a(x'_i)$, whereupon $\theta_a(x'_i)$ will be small
too, and the rotating frame of reference will coincide with the intrinsic one.
The definition of the intrinsic system will be addressed again below,
now the quantum constraint conditions come into consideration.
As the work in Ref.~\cite{LM98}, the present one makes use of approximate 
constraints, but because of their importance, provides more rigorous 
substantiation and shows what results follow from them.

Correct, but with a need to overcome great difficulties, methods 
for taking into account quantum constraints for the nuclear rotation 
were proposed in Ref.~\cite{Bes86} (and references therein) 
on the basis of Faddeev-Popov functional integral and
in Ref.~\cite{Bes91} on the basis of BRST-symmetry.
We will use the method of Ref.~\cite{Bes86} and confine 
ourselves to its brief statement, indicating only suggested changes to it.

In classical mechanics, constraints can be introduced as follows.
If the Lagrangian in the laboratory system is assumed independent of additional
variables $\zeta_a$, there arise constraints for momenta,
canonically conjugate to these variables:
$\varPi_a \equiv \partial L/\partial \dot{\zeta}_a = 0$.
The constraints are introduced into the Hamiltonian as an additional term
with Lagrange factors, while equations $\zeta_a = 0$ can be chosen
as gauge conditions.

In quantum mechanics, when the vacuum transition amplitude 
for a nucleus is written in terms of a path integral, 
the additional variables in the laboratory frame are introduced 
into the path integral with the multiplier
\begin{equation} \label{1=}
1 = \int \prod_{ab} \delta(\varPi_a ) \; \delta(\zeta_b) \;
D\varPi_a D\zeta_b  \:,
\end{equation}
where $\delta(\ldots)$ is delta function. Computation of the resulting 
functional integral can be replaced (see \cite{Bes86} and references therein)
with solving the eigenvalue problem for the effective Hamiltonian
\begin{equation} \label{Heff}
H_{\mbox{eff}} = \lim_{{\mathcal D} \to 0}
\left\{H + \frac{1}{2{\mathcal D}} \sum_a \varPi_a^2(\zeta) +
\frac{1}{2{\mathcal A}} \sum_a \zeta_a^2\right\} \;.
\end{equation}
Here $H$ is the nucleus Hamiltonian in the laboratory system, 
$\mathcal A$ is an arbitrary constant, having no effect 
on the final expressions;
$\varPi_a(\zeta) \equiv -i \partial / \partial \zeta_a $
are conjugate to the angular variables.

In the Hamiltonian $H_{\mbox{eff}}$, the spurious motion, described
by the additional variables $\zeta_a$, is separated from the real one 
and represents the harmonic oscillator motion. We will be interested
only in those eigenstates of the total Hamiltonian (\ref{Heff}),
which correspond to the ground state of the ghost motion.
At ${\mathcal D} \to 0$, the states are separated by infinite energy 
from all states, where the spurious oscillator is excited. 
After transformation to the IF variables,
the nucleonic and supplementary variables in the Hamiltonian
(and in wave functions) are not separated any more, but the limit
${\mathcal D} \to 0$ ensures the fixing of the redundant degrees of freedom
and the absence of spurious motion. For the ground state of the oscillator,
we have $\langle \zeta^2_a \rangle = 1/2\sqrt{{\mathcal A}/{\mathcal D}}$
and, with $\mathcal A$ fixed, the relationship
$\langle \zeta^2_a \rangle \to \infty$ at ${\mathcal D} \to 0$ is valid,
which makes it impossible to ensure smallness of $\theta_a$. Yet,
the freedom in selecting $\mathcal A$ allows to take
it such that ${\mathcal A} \to 0$ at ${\mathcal D} \to 0$ and the ratio 
${\mathcal A}/{\mathcal D}$ were small enough.
This ensures sufficient smallness of $\langle \zeta^2_a \rangle$,
and therefore, after the ${\mathcal U}$-transformation, smallness of 
$\langle \theta^2 \rangle$. Note that one condition ${\mathcal A} \to 0$
suffices for the approximation used below, whereas the limit 
${\mathcal D} \to 0$ is introduced here to trace a parallel
with Ref.~\cite{Bes86}.

Since at $\zeta \to 0$, the matrix $b^{-1}_{ab}(\zeta)$, inverse 
with respect to $b_{ab}(\zeta)$, reduces to Kronecker $\delta_{ab}$ symbol 
and since we will use only those states of the Hamiltonian $H_{\mbox{eff}}$,
where the spurious oscillator is in the ground state, one can replace
$\varPi_{\alpha}(\zeta)\; \varPi_{\alpha}(\zeta)$
in Eq.~(\ref{Heff}) by the operator
$b^{-1}_{\alpha\beta} \; b^{-1}_{\alpha\gamma} \; \varPi_{\beta} \; \varPi_{\gamma} \;$.
Validity of this change is especially clearly seen in the path integral,
corresponding to the Hamiltonian (\ref{Heff}), because of the presence of
$\delta(\zeta_a)$ in the integral. Considering the above and the commutation relation
$[\varPi_{\beta}\;, b^{-1}_{\alpha\gamma}] \to$ $\varepsilon_{\alpha\beta\gamma} \,i/2$
at $\zeta_a \to 0$, one can finally replace the operator 
$\varPi_{\alpha} \varPi_{\alpha}$ in Eq.~(\ref{Heff}) by the operator
$I_{\alpha} I_{\alpha}$. Here $\varepsilon_{\alpha\beta\gamma}$ 
is the unit antisymmetric tensor and $I_{\alpha}(\vartheta) \equiv$
$D^1_{\alpha\beta}(\vartheta^{-1}) \; \grave{I}_\beta$ is
the projection of the angular momentum of a rigid body  to the intrinsic axes,
$\grave{I}_\beta = b_{\beta\gamma}\varPi_{\gamma}$.

After the transformation to IF variables $\{ x_i, \zeta_a \}$ $\to$ 
$\{x'_i, \vartheta_a \}$, operators, involved in Eq.~(\ref{Heff}),
have the form
$I_a(\zeta) = I_a(\vartheta) - J_a(x'_1,\ldots ,x'_A)$, \ 
$\zeta_a = -\theta_a(x'_1,\ldots ,x'_A)$, and the effective Hamiltonian 
itself is described by the expression
\begin{equation} \label{tildeHeff}
\widetilde{H}_{\mbox{eff}} = \lim_{{\mathcal D,A} \to 0}
\left\{\widetilde{H} + \frac{1}{2{\mathcal D}} \sum_a (I_a-J_a)^2 +
\frac{1}{2{\mathcal A}} \sum_a \theta_a^2\right\} \;
\end{equation}
Note that the effective Hamiltonian in the approach \cite{Bes86} resembles
Eq.~(\ref{tildeHeff}), but contains the Hamiltonian $H$ instead of $\widetilde{H}$.
Therefore the description of a nucleus in IF is for the most part accomplished
only due to the constraints and the precision of the method is strongly limited
with accuracy of accounting for constraints. In the suggested approach,
as will be seen below, good results  are produced by accounting for constraints 
even in the lowest approximation.

The method of Ref.~\cite{Bes86}, in its nowaday formulation, is practically
inapplicable to self-consistent description of nuclei, as it requires an exact
compensation of divergent (in the limit ${\mathcal D} \to 0$) terms
in every order of the perturbation theory. Used below is another,
more primitive method of accounting for constraints,
which nonetheless contains the most important effects,
described by the Hamiltonian (\ref{tildeHeff}).
Instead of the Hamiltonian $\widetilde{H}_{\mbox{eff}}$, we will use
$\widetilde{H}$, but with account of two important properties of eigenfunctions
$\widetilde{\Psi}_I$ of the Hamiltonian $\widetilde{H}_{\mbox{eff}}$.
Firstly, the expectation value
$\langle \widetilde{\Psi}_I | \theta^2_a| \widetilde{\Psi}_I\rangle$
should be small, and, hence, ME of operators $\theta_a$
should be small, and, secondly, (see \cite{ML94,LM95}),
\begin{equation} \label{<IJ>}
\langle \widetilde{\Psi}_I | \mathbf{I}\mathbf{J}| \widetilde{\Psi}_I\rangle = I(I+1) \;.
\end{equation}

The second of the conditions is simpler, so we start discussing it first.
Correctness of (\ref{<IJ>}) can be checked by transforming its left-hand side from
intrinsic to laboratory variables. This yields the relationship for operators:
\begin{equation} \label{U-1IJU}
{\mathcal U}^{-1}I_\alpha J_\alpha {\mathcal U} = J_\beta J_\beta -
J_\beta \, D^1_{\gamma \beta}(\theta^{-1}(x_i)) \, I_\gamma (\zeta)\, .
\end{equation}
Upon averaging over eigenfunction $\Psi_I$ of the Hamiltonian $H_{\mbox{eff}}\,$
(\ref{Heff}), the second term in the right-hand side (\ref{U-1IJU}) yields 0,
since $\Psi_I$ is separable in functions, dependent on nucleonic and additional
variables, with the ground-state function of the auxiliary oscillator being real
and satisfying the equation $\langle I_\gamma \rangle = 0$.
In the end, the expectation value of (\ref{U-1IJU}) is equal to the right-hand
side of (\ref{<IJ>}).

The Hamiltonian (\ref{tildeHeff}) and operator $ \mathbf{I}\mathbf{J}$,
involved in the equation (\ref{<IJ>}), act in the unified space of nucleonic and
angular variables,  that greatly complicates application of traditional methods
of the many-body theory. To simplify the problem, the ME of operators $I_a$,
acting in the space of collective angles (so called geometrical factors), can be
found in different ways depending on the state spins.
The ME for band-head states and close to them rotational levels
with $I \simeq K$, can be taken into account exactly.
Here, parts of the Hamiltonian, containing geometrical factors,
should be diagonalized in a basis of low dimensionality, which is built of states
with different values of the quantum number $K$. The wave functions of intrinsic
motion in these states are weakly dependent on $I$, and hence,
little differ from wave functions of zero-order approximation,
i.~e. with rotation ignored, so iterative processes in self-consistent calculations
with account of quasiparticle-rotation interaction will converge rather fast
(as compared to states with large $I$). Another, more simple and in many cases
preferable, way of treating the terms with the operators $I_a$ directly
in the unified space of nucleonic and angular variables is to implement
a semi-perturbative approximation, considered in Sec.\ref{sect4}. 
For the band-head states and close to them rotational levels, the operator $F$,
introduced in this section, is very small that insure the high accuracy
of the approximation.

Good accuracy for rotational states that are not band-head states is achieved
with the standard SCM approximation; to apply it, one should get rid of collective
angles $\vartheta$ and make use of approximate equations:
\begin{equation} \label{I_x}
I_x \simeq \sqrt{I(I+1)-\langle J^2_z \rangle} \:, \qquad
I_y \simeq 0 \;.
\end{equation}
This is carried out by multiplying the corresponding equations, e. g.,
those for eigenvalues of the Hamiltonian $ \widetilde{H}$
(or $ \widetilde{H}_{\mbox{eff}}$), by $D$-functions,
then integrating the equations over $\vartheta$ and summing over $K$
with subsequent use of Eqs.~ (\ref{I_x})
(see Ref.~\cite{LM91}).
Note that such approximation proves in many cases to be acceptable
(when calculating energies) even for band-head states. Of interest in
the present work is mostly the qualitative aspect of the problem,
so we confine ourselves to the SCM approximation (\ref{I_x}). 
Then the quantum constraint (\ref{<IJ>}) reduces to the ordinary SCM condition 
on the nuclear wave function $|\Phi_I\rangle$, dependent on nucleonic variables only:
\begin{equation} \label{J_x}
\langle J_x \rangle = \sqrt{I(I+1)-\langle J^2_z \rangle} \,.
\end{equation}

Consider now the first approximate constraint condition: smallness
of the ME of operators $\theta_a$. On the one hand, the smallness arises
as a consequence of the limiting process in the
equation (\ref{tildeHeff}), resulting in small expectation value
$\langle \theta^2 \rangle$ (and similar averages for higher powers of $\theta$).
This allows simplifying the initial expressions for the Mikhajlov Hamiltonian and
commutation relations, and more precisely, their contribution
to the variation functional. The functional, built on their basis,
represents a series in powers of ME $\theta_a$ and in powers of products
of operators $\theta_a\,(I_b-J_b)$. From the conditions of $\theta$ smallness
it follows that among terms, differing in the powers of $\theta$, only terms
of the lowest power contribute to the functional.

On the other hand, to ensure that the condition holds, it is necessary
to retain in the variation only the terms of the lowest power in $\theta$,
i.~e. use the expressions (\ref{simple_H}) and (\ref{simple_commut})
in the functional. Indeed, the ME of operators $\theta$ are determined
by minimization of the functional $\langle \widetilde{H} \rangle$ 
on the hypersurface, defined by the equation (\ref{commutation}). 
The varying functional is a multidimensional polynomial of a very high degree 
in the ME and can have very many minimums. For the condition of $\theta$ 
smallness to hold, the functional minimum should be found,
which position is close to zero values of the ME. To find it all polynomial 
terms of power higher than two should be discarded.

Smallness of $\theta$ does not imply smallness of $\theta_a\,(I_b-J_b)$, as:
i) by virtue of Eq.~(\ref{<J,theta>}), coherent sum of $\theta_a\,J_a$
over states of all nucleons is of order 1, and ii) the value
$I_x$ increases with increasing nuclear spin. However, strong
dependence of terms of high power in $\theta\,(I-J)$ on spin
$I$ is compensated by the same dependence on $\langle J_x
\rangle$ with the account of Eq.~ (\ref{J_x}). Because of
the presence of commutators in Eq.~(\ref{simple_H}), each
multiplication by $\theta\,(I-J)$ is linked with additional
coherent summation only in terms of even power in this factor.
Here such summations appear in expectation values $\langle J_a
J_a \rangle$. But the number of couples $J_a J_a$ is half of
number of the operators $\theta$, therefore every additional
odd power by $\theta\ J$ gives an additional small factor,
which can be evaluated as one-particle contribution to $\langle
[J_a,i\theta_a] \rangle$ (see the next section). Thus, only
three first terms can be retained in the series (\ref{simple_H})
for $\widetilde{H}$.

Finding the operators $\theta_a$ by the variational method, one, by that very act,
define the optimal IF (OIF). In our approximation, this is the frame of reference,
where a nucleus is described to the best advantage 
by a single-quasiparticle wave function.
Similarly, the OIF can be defined for a more general form of wave functions
(the OIF in the random phase approximation for even-even nuclei is
briefly discussed in Ref.~\cite{LM98}); this is, however, beyond the scope
of the present work. The general scheme of the iterative computation within 
the given approximation is as follows. The ME of operators $\theta_a$ 
are obtained by minimizing the expectation value $\langle \widetilde{H} \rangle$
under the supplementary condition
\begin{equation} \label{<J,theta>}
\langle [J_a,i\theta_a] \rangle =1 \,.
\end{equation}
The initial stage uses eigenfunctions of the SCM Routhian
$H_\omega \equiv H-\omega J_x$ with the angular frequency $\omega$
determined by Eq.~(\ref{J_x}). Thus, the variational functional has the form:
\begin{equation}  \label{F}
\mathcal{F} =  \langle  H \rangle +
\sum_{a=x,y} \left\{ - \langle [ H,i\theta_a ] \Delta J_a \rangle \:
+\: \frac{\langle (\Delta J_a)^2\rangle}{2}
\langle [ [ H,i\theta_a ],i\theta_a ] \rangle
+ \eta_a \left( 1-\langle [J_a,i\theta_b] \rangle \right) \right\},
\end{equation}
where $\eta_a$ are Lagrange multipliers, 
$\Delta J_a  =  J_a  - \langle J_a\rangle $.
After determining $\theta_a$, the Hamiltonian
$\widetilde{H}$ is constructed, and eigenfunctions of the
Hamiltonian $\widetilde{H}_\omega \equiv \widetilde{H} - \omega
J_x$ are sought with the condition (\ref{J_x}). These new
functions are used to find more correct $\theta_a$, and so on,
until the convergence of iterative process.
\section{Many-particle operators of nuclear orientation angles \label{sect3}}
%
The operators $\theta_a$ are many-particle ones and so, 
in the representation of creation and
annihilation operators $a^{\dagger}$, $a$, they have the form
\begin{equation} \label{Np-theta}
\theta = \frac{1}{N!} \sum \theta_{12\dots N,\, 1'2'\ldots N'} \,
a^{\dagger}_1 a^{\dagger}_2 \ldots  a^{\dagger}_N
a_{N'} \ldots a_{2'} a_{1'} \,,
\end{equation}
where the ME $\theta_{12\dots N,\, 1'2'\ldots N'}$ 
of the operator $\theta_a$ are written in the
antisymmetrized form, index $a$ ($a=x,y$) is omitted. Direct calculation 
of the ME by variational method is hardly feasible: 
$N$-particle operator in the basis of dimensionality $M$
requires variation of $M^{2N}$ parameters. 
Yet, the situation changes radically, if one takes into account that, 
with proper selection of the basis, only some combinations of the ME
actually contribute to the observable quantities.

A convenient choice of the initial basis in (\ref{Np-theta}) is the canonical basis
(see \cite{RS,Dobaczewski_1996}), obtained by the HFB method for Hamiltonian $H$
(hereinafter we consider the Hamiltonian with effective forces), or,
for the sake of simplicity, single-particle functions,
obtained by the HF method for the same Hamiltonian. Then, in representation
of quasiparticle operators $\alpha^{\dagger}$, $\alpha$, obtained
by the HFB method for the SCM Routhian $H_\omega$
(or Hamiltonian $\widetilde{H}_\omega$),
operator $\theta$ has the form
$\theta = \theta^{11} + \theta^{20+02} + \theta^{40+04+22+31} + \ldots \;$.
Hereinafter, for any operator $A$ brought to normal form, symbol $A^{ij}$
denotes its part, containing $i$ operators $\alpha^{\dagger}$ and $j$
operators $\alpha$. Applications of similar denotation in terms of other
creation operators will be always accompanied by the pointing
to which operators they pertain.

To derive variational equations for the ME of operators $\theta_a$, 
it is necessary first to estimate order of smallness
of their different components and use the estimates 
to simplify the functional under variation.
We begin consideration with $\theta^{11+20}_a$. Comparison of quantities
$\langle [J_a,i\theta_a] \rangle =1$ and
$\langle \Delta J^2_a \rangle $
shows that $\theta^{11+20}_a \sim J^{11+20}_a/2\langle \Delta J^2_a \rangle$
on the average, and, since $\langle \Delta J^2_a \rangle \sim 50$
for deformed nuclei, one can conclude that $\theta^{11+20}_a$  is two orders
of magnitude less than $J^{11+20}_a$.
In the Hamiltonian (\ref{simple_H}), only each even power of $\theta J$
is linked with coherent summation over states of all particles,
whereas each odd power is not accompanied by such summation due to the presence
of commutators and selection rules for the ME of operators $\theta$ and $J$.
Therefore, terms of power higher than two in $\theta^{11+20}_a J^{11+20}_a$
in (\ref{simple_H}) can be neglected.

To simplify further, consider the structure of operators $\theta^{ij}$.
In each $\theta^{ij}$, multipliers before normal products of the quasiparticle
creation and annihilation operators have the form of foldings of the ME
$\theta_{12\dots N,\, 1'2'\ldots N'}$ with density matrices
$\rho_{lm}=\langle a^\dagger_m a_l \rangle$ and pair tensors
$\kappa_{lm}=\langle a_m a_l \rangle$,
$\kappa^*_{lm}=\langle a^\dagger_l a^\dagger_m \rangle$;
the summands of the operator $\theta^{ij}$ contain products of different
numbers of matrices $\rho$ and $\kappa$.
For example, the structure of $\theta^{11+20}$ is:
\bea
&&\theta^{11+20} = \theta^{11+20}(a^{\dagger} a) +
\theta^{11+20}(a^{\dagger} a^{\dagger} + aa) \,,\label{theta} \\
&&\theta^{11+20}(a^{\dagger} a) =
\sum_{pp'}
\left[ \:f_0(N) \:\widehat{\theta \rho^{N-1}} +
f_1(N) \:\widehat{\theta \kappa\rho^{N-2}} + \, \ldots \,
+ f_{N-2}(N) \:\widehat{\theta \kappa^{N-2}} \,
\right]_{pp'} \nonumber \\
&& \phantom{\theta^{11+20}(a^{\dagger} a) = \sum _{pp'}w}
\times :a^{\dagger}_p \,a_{p'}: \,,\label{theta-1} 
\eea
\bea
&& \theta^{11+20}(a^{\dagger} a^{\dagger} + aa) =
\frac{1}{2} \sum_{pp'}
\left\{ \left[ \: g_1(N) \:\widehat{\theta \kappa\rho^{N-2}}+\, \ldots \,
+ g_{N-2}(N) \:\widehat{\theta \kappa^{N-2}} \,
\right]_{pp'}:a^{\dagger}_p \,a^{\dagger}_{p'}: \right. \nonumber \\
&& \phantom{\theta^{11+20}(a^{\dagger} a^{\dagger} + aa)= \sum _{pp'}}
+\left.\left[ \: g_1(N) \:\widehat{\theta \kappa\rho^{N-2}}+\, \ldots \,
+ g_{N-2}(N) \:\widehat{\theta \kappa^{N-2}} \,
\right]^*_{pp'}  :a_{p'} \,a_p:  \right\}, \label{theta-2}\\
&&\left[ \widehat{\theta \rho^{N-1}}\right]_{pp'} \equiv
\sum_ {2\dots N,\, 2'\ldots N'} \theta_{p2\dots N,\, p'2'\ldots N'}\,
\rho_{2'2} \, \ldots \, \rho_{N'N}\,,\label{theta-3} \\
&& \left[\widehat{\theta \kappa\rho^{N-2}}\right]_{pp'} \equiv
\sum_ {2\dots N,\, 2'\ldots N'} \theta_{pp'3\dots N,\, 22'3'\ldots N'}\,
\kappa_{22'} \, \rho_{3'3}\, \ldots \, \rho_{N'N}\,.\label{theta-4}
\eea
Here expressions, like $:a_{p'}a_p:$, denote normal product of
operators, obtained after the~Bogoliubov transformation, $f_k(N)$
and $g_k(N)$ are combinatorial factors, dependent on $N$, in~particular: 
$f_1(N)=N$, $g_1(N)=N(N-1)/2$. Of similar structure are $\theta^{ij}$ 
with large values $i+j$.

The less operators $\alpha^{\dagger}$ and $\alpha$ are contained in component
$\theta^{ij}$, i.~e. the less $i+j$, the more coherent summations (foldings
$\theta_{12\dots N,\, 1'2'\ldots N'}$ with $\rho_{lm}$ or with $\kappa_{lm}$)
are contained in the multiplier of operator
$\alpha^{\dagger}\ldots \alpha^{\dagger}\alpha \ldots\alpha$.
The convolution $\widehat{\theta\rho}$ contains coherent summation
over all filled nucleonic states. Such sum can be estimated by order
of magnitude as one term, multiplied by the number of nucleons $A$.
This folding proves to be much greater in magnitude than
the folding $\widehat{\theta \kappa}$, since the latter contains
summation only over single-particle states in the vicinity of the Fermi surface
where occupation probabilities are smeared out by the pairing correlations.
Note that with no pairing correlations, one has: $\widehat{\theta \kappa} =0$.
Similar, but approximate relation, is useful to simplify the description
of highspin states and also other nuclear states with slight pairing.

However, multipliers of operator 
$\alpha^{\dagger}\ldots \alpha^{\dagger}\alpha \ldots\alpha$ 
contain in addition to the foldings also factors $f_k(N)$ or $g_k(N)$ that
describe the number of ways to obtain the foldings.
If $N=A$, i.~e. the operator $\theta$ is $A$-particle one 
and contribution of each nucleon to the operator depends on contributions 
of all other nucleons of the nucleus,
then the combinatorial factors outside $\kappa_{lm}$ are, in many cases,
greater than the multipliers of $\rho_{lm}$. Nonetheless, 
by virtue of the condition that $\theta$ are small enough, 
the contribution of a nucleon to angle of deviation of nuclear
axes from the intrinsic frame is dependent only on a small number of other nucleons
and, hence, $N<A$ (as a nice illustration, one can consider an expression 
for a small angle of orientation of the tensor-of-inertia axes for $A$-particles).
For example, in case of the two-particle operator $\theta$, all factors
$f_k(N)$, $g_k(N)$ are of the same order of magnitude,
and therefore terms with $\kappa_{lm}$ are negligible. There is an
analogy here with components of the two-particle Hamiltonian: its components,
containing $\kappa_{lm}$, contribute very little to the total energy of the nucleus
as against those with $\rho_{lm}$ (their contributions to excitation energies
of the nucleus are comparable, but excitation energies of themselves
are much less than binding energy). Yet, contributions of $\theta$ exhibit
an important difference: they are themselves of some order
of smallness, and the corrections that arise in taking account of components $\theta$,
containing $\kappa_{lm}$, are corrections to corrections. In a more general case,
even if $N\sim A/4$, sums with $\rho_{lm}$ prove to be much greater than those
with $\kappa_{lm}$ (see the preceding paragraph). Such estimate is
of qualitative nature. In view of importance of this problem,
we will derive a quantitative estimate in another independent way.

To evaluate $\theta^{40+31+22}$, we differentiate the variational functional with
respect to the ME of the operators $\theta^{40}$. Since, by virtue of their structure,
ME of $\theta^{40}$ are on the average not greater than ME of $\theta^{20+11}$,
and the values of ME of $\theta^{31}$ and $\theta^{22}$ are on the average
of the same order of magnitude as the ME of $\theta^{40}$,
it suffices to retain in the functional the terms of power not higher than two in
these operators. At that (for estimation), there is no need to derive
all expressions in detail: it suffices to analyze only the structure
of the resulting linear equation and take into account the validity,
on the average, of the following estimate for ME:
$\partial\theta^{31} / \partial\theta^{40} \sim 1$ .
Then we get that 
$\theta^{40} \sim \theta^{20}J^{11}J^{20} / \langle \Delta J^2_a \rangle $. 
By using the above estimate for $\theta^{20}$, one can easily make sure
that $\theta^{40} << \theta^{20}$ and, hence, operators 
$\theta^{ij}$ with $i+j=4$ can be neglected. 
A more thorough analysis confirms validity of the conclusion even in cases,
when the average of the transformed Hamiltonian has the number 
of summations over single-particle indices for some terms, containing $\theta^{40}$, 
in excess of that in similar terms, containing $\theta^{20}$.

The similar way can be used to derive estimate for $\theta^{60}$ and show that $\theta^{60}
<< \theta^{40}$. Extending the considerations further, we get finally that operators
$\theta^{ij}$ with $i+j>2$ can be neglected.

Thus, contribution to observable quantities comes from the operators
\begin{eqnarray}
&&\theta^{11+20}(a^{\dagger} a) =
\sum \theta^{(\rho)}_{pp'}\,:a^{\dagger}_p \,a_{p'}: \,,\label{theta-5} \\
&&\theta^{11+20}(a^{\dagger} a^{\dagger} + aa) = \frac{1}{2}
\sum \left[ \:\theta^{(\kappa)}_{pp'}\,:a^{\dagger}_p \,a^{\dagger}_{p'}: \;+
\;\: \theta^{(\kappa)*}_{pp'}\,:a_{p'} \,a_p: \, \right] \,,\label{theta-6}\\
&&\theta^{(\rho)}_{pp'} \equiv N\sum_ {2\dots N,\, 2'\ldots N'}
\theta_{p2\dots N,\, p'2'\ldots N'}\,
\rho_{2'2} \, \ldots \, \rho_{N'N}\,,\label{theta-7} \\
&&\theta^{(\kappa)}_{pp'} \equiv \frac{N(N-1)}{2}
\sum_{2\dots N,\, 2'\ldots N'} \theta_{pp'3\dots N,\, 22'3'\ldots N'}\,
\kappa_{22'} \, \rho_{3'3}\, \ldots \, \rho_{N'N}\,,\label{theta-8}
\end{eqnarray}
where multipliers outside normal products contain maximum number
of density matrices.

In principle, all coefficients $\theta^{(\rho)}_{pp'}$, $\theta^{(\kappa)}_{pp'}$
can be found by variational method. This does not complicate calculations much
as against the approximation, when only operators (\ref{theta-5}) are taken
into account. However, the present work will make use of a certain smallness
of $\theta^{(\kappa)}$ as compared with $\theta^{(\rho)}$ and neglect
the operator (\ref{theta-8}). To make sure of smallness of $\theta^{(\kappa)}$,
compare expressions (\ref{theta-7}) and (\ref{theta-8}).
Although the coefficients $\theta^{(\kappa)}_{pp'}$ contain an additional factor
$(N-1)/2$, the factor is compensated by smallness of the convolution
$\widehat{\theta \kappa}$ as against the $\widehat{\theta \rho} \,$:
the former can be approximately estimated as sum over all filled states,
and the latter --- as sum over states near the Fermi surface,
participating in pair correlations.
But $\theta^{(\kappa)}_{pp'}$ has besides an additional smallness,
arising for the following reasons. Integration over coordinates in the ME
$\theta_{pp'3\dots N,\, 22'3'\ldots N'}$ binds single-particle wave functions
$\phi_p(x_1)$ with $\phi_2(x_1)$, and functions $\phi_{p'}(x_2)$ with
$\phi_{2'}(x_2)$. Contribution to expression (\ref{theta-8})
is made only by states $|2\rangle$, $|2'\rangle$ near Fermi level,
otherwise $\kappa_{22'} = 0$. But then, for states $|p\rangle$, $|p'\rangle$
far from Fermi level, overlaps of functions $\phi_p(x_1)$ with $\phi_2(x_1)$
and $\phi_{p'}(x_2)$ with $\phi_{2'}(x_2)$ are small and, hence,
$\theta^{(k)}_{pp'}$ are small. Further, the states $|p\rangle$ and $|2\rangle$
must have identical isospin projections and also parity (the latter holds because
of positive parity of the operator $\theta$).
With decreasing pair correlations, $\kappa_{22'} \to 0$ and, hence,
$\theta^{(\kappa)} \to 0$. Under slight mixing of states in the quantum
number $K$, defined by the equation $J_z|p\rangle = K_p|p\rangle$,
single-particle states in the ME $\theta_{pp'3\dots N,\,22'3'\ldots N'}$
should satisfy approximate selection rules: $K_p-K_2=\pm 1$,
$K_{p'}-K_{2'}=0$. This follows from the variational equations for $\theta$,
which show that operators $\theta^{11+20}$ and $J^{11+20}$ have
identical selection rules by $K$, and this is also demonstrably illustrated
by the example of rotation angle of tensor-of-inertia principal axes
at small values of the angle,
when the function '$\arctan$' that is involved in the definition of the angle can be
replaced by its argument and the resulting expression can be rendered
in terms of spherical functions. In the model with monopole pairing,
pair tensor is diagonal: $\kappa_{22'}=\delta_{2'\bar{2}} \,\kappa_{2\bar{2}}$
(the bar under index denotes time conjugation).
Hence it follows that contribution to expression (\ref{theta-8}) is made only
by two states $|2\rangle$, approximately conforming to requirement
$K_p-K_2=\pm 1$, with the contribution suppressed by the multiplier
$\kappa_{2\bar{2}} \simeq u_2v_2$, consisting of Bogoliubov transformation
coefficients. No such restrictions exist for quantities $\theta^{(\rho)}_{pp'}$,
so, to simplify, one can neglect the operator
$\theta^{11+20}(a^{\dagger} a^{\dagger} + aa)$. It can be mentioned that 
the K-mixing does not influence essentially to the conclusion on
the smallness of $\theta^{(\kappa)}_{pp'}$ because, for a higher mixing, 
the abovementioned arguments may be applied to each component and 
because the decreasing of the pairing, caused by the mixing,
will diminish $\theta^{(\kappa)}_{pp'}$.

The quantities $\theta^{(\rho)}_{pp'}$ are dependent on density matrices and
so can change with $u$, $v$ changing.
However, within the framework of the semi-perturbative approximation
(at not too big nuclear spins), used in the next section, the changes
can be shown negligible. The operator (\ref{theta-5}) can in such case
be considered as some analog of a single-particle operator,
i.~e. as a~"quasi-singleparticle" one:
$ \sum_{pp'} \langle p|\theta_a|p'\rangle\,a^{\dagger}_p a_{p'}$,
where $\langle p|\theta_a|p'\rangle\ = (\theta_a)^{(\rho)}_{pp'}$ and $a=x,y$.
In the general case, when the changes are
essential, the difficulties, related to such dependence of
$\theta^{(\rho)}_{pp'}$, are easily overcome, if the Hamiltonian $\widetilde{H}$
at each iteration is expressed in terms of operators $\alpha$,
obtained at the preceding iteration. To clarify this, consider, for example,
first two steps of the iterative procedure.

At the first step, after the transformation
$a^\dagger_p =u_{p\mu} \alpha^\dagger_\mu +v_{p\mu} \alpha_\mu$,
one obtains the coefficients $u,v$ from the HFB equations 
for the Routhian $H_\omega$ and then solves the variational equations 
for $\theta^{(\rho)}_{pp'}$.
At the second step, one uses the same Bogoliubov transformation to express
the Hamiltonian $\widetilde{H}_\omega  = \widetilde{H} - \omega J_x$
in terms of $\alpha^\dagger_\mu$, $\alpha_\mu$ and substitutes the obtained
values $\theta^{(\rho)}_{pp'}$. After that, a new transformation
$\alpha^\dagger_\mu ={\tilde u}_{\mu \nu}{\tilde \alpha}^\dagger_\nu +
{\tilde v}_{\mu\nu} {\tilde \alpha}_\nu$ of the Hamiltonian
$\widetilde{H}_\omega $ is performed with the fixed ME
$\theta^{(\rho)}_{pp'}$ in Eq.~(\ref{theta-5}). These 
${\tilde u}$, ${\tilde v}\,$-coefficients are determined by HFB equations
for $\widetilde{H}_\omega $. After the transformation
$a^\dagger_p ={\tilde U}_{p \nu}{\tilde \alpha}^\dagger_\nu +
{\tilde V}_{p\nu} {\tilde \alpha}_\nu$, with the obvious expressions of
${\tilde U}_{p \nu}$, ${\tilde V}_{p\nu} $ via $u_{p\mu}$, $v_{p\mu}$
and ${\tilde u}_{\mu \nu}$, ${\tilde v}_{\mu\nu}$, one obtains the functional
$\mathcal{F}$ in terms of  new ME $\theta^{(\rho)}_{pp'}$,
which are again determined by the variational equations,
but with the new coefficients ${\tilde U}_{p \nu}$, ${\tilde V}_{p\nu}$.
Thus, such iterative procedure effectively takes into account that
the values $\theta^{(\rho)}_{pp'}$ contain in every step not only new
ME $\theta_{p2\dots N,\, p'2'\ldots N'}$, but also new density matrices.

To compare the present method with the approximate AMP (AAMP), based
on the Kamlah expansion, one can consider some explicit forms
for $\theta$ ME and $\langle \widetilde{H} \rangle$. Such expressions are given
in Ref.~\cite{LM98} for ground-state bands of even-even nuclei,
but similar formulas may be obtained for odd nuclei and excited bands
of even ones. For example, consider the functional $\mathcal{F}$
(Eq.~(\ref{F})) for the state vector
$|s\rangle \equiv \alpha^\dagger_s|0\rangle$ with $u,v$ coefficients,
satisfying the HFB equations for the Routhian $H_\omega$ (the HFB
approximation is not crucial here). Variation of $\mathcal{F}$ with
the approximate relations $\Delta [J_a,i\theta_b] \simeq 0$,
$\Delta [H,i\theta_a ] \approx \Delta [H_\omega,i\theta_a ]$ yields the ME:
\begin{eqnarray}
&& \langle 0| \alpha_t\, i \theta_a\, \alpha^+_s\, |0\rangle =
\frac{1}{2\langle \Delta J^2_a \rangle}\,\langle 0| \alpha_t\, J_a\,
\alpha^+_s\, |0\rangle \:, \label{explicit_theta_1} \\
&& \langle 0| \alpha_s\,\alpha_q\,\alpha_p\,i \theta_a\,
\alpha^+_s\, |0\rangle =
\frac{1}{2\langle \Delta J^2_a \rangle}\,\langle 0|
\alpha_s\,\alpha_q\,\alpha_p\, J_a\, \alpha^+_s\, |0\rangle \,.
\label{explicit_theta_2}
\end{eqnarray}
There are the same equations for the Hermitian conjugated ME, but 
the minus sign appears in the right-hand side of the equations. 
After substituting of these ME, $\langle \widetilde{H} \rangle$
takes the form of the well-known AAMP expression for the
mean value of the nucleus Hamiltonian (see, the review \cite{Ring_1985}).
Some more complex consideration makes it possible to obtain, also,
the AAMP expression without exploiting the Eq.~(\ref{I_x}).
But the determination of the quantum recoil
corrections to wave functions in OIF differs significantly from that of AAMP.
Indeed, in the next and following iterations of the AAMP method,
$u,v\,$-coefficients are obtained by a minimization of the same expression for
the energy (now, with the recoil term), while in the OIF description, $u,v$
are determined by HFB equations for the Hamiltonian $\widetilde{H}_\omega$,
that is, by the minimization of $\langle \widetilde{H}_\omega \rangle$.
The last functional would coincide with the AAMP one,
if Eq.~(\ref{explicit_theta_1},\ref{explicit_theta_2}) were valid for all variations
of  the vector $|s\rangle$. But any variation of $|s\rangle$ changes the form of
$\theta$ ME. For example,  a particular variation
$|s+\delta s\rangle  = \sum_{t'} C_{t'}\,| t'\rangle $
gives ME $\langle t| i \theta_a|s+\delta s\rangle $ =
$\sum_{t'} C_{t'}\,\langle t| i \theta_a| t'\rangle $, where all ME with $t' \ne s$
are quit different from Eq.~(\ref{explicit_theta_1}, \ref{explicit_theta_2}). To obtain
the ME with $t' \ne s$, one has to solve the system of
Eq.~(\ref{explicit_theta_1}, \ref{explicit_theta_2}) with respect to $\theta^{(\rho)}_{pp'}$
and then, with the help of these values, calculate $\langle t| i \theta_a| t'\rangle $.
The difference between AAMP and OIF arises, probably, because AAMP is not justified
in the cases of strong band mixing. More over, it should be mentioned
that the variational determination of $\theta^{(\rho)}_{pp'}$
is much more preferable than the determination of the values
$\langle 0| \alpha_t\, i \theta_a\, \alpha^+_s\, |0\rangle$,
$\langle 0| \alpha_s\,\alpha_q\,\alpha_p\,i \theta_a\, \alpha^+_s\, |0\rangle$:
see the next section.
\section{Detailing of the method and results of calculations in a simple model \label{sect4}}
%
As noted in Introduction, main features of the proposed method are
conveniently analyzed within the framework of a relatively simple,
but realistic model, rather than in full-scale microscopic calculations.
To this end, the present work employs a semi-perturbative approximation, combining
exact diagonalization with perturbation theory. Despite its simplicity, such approach
has its advantages: the approximation
i)~allows taking account of the blocking effect for states, mixed by
the quasiparticle-rotation interaction,
ii)~ensures orthogonality of all resulting states, which poses a certain problem
in the presence of terms with Lagrange factors $\eta_a$ and $\omega$
in the functional,
iii)~yields good convergence of iterative procedures and
visual revealing of roles of various effects,
iv)~gives the possibility to treat exactly the geometrical factors, mentioned
in Sec.~\ref{sect2}.
Detailed statement of the semi-perturbative approximation (with exact account
of the geometric factors, but without the recoil operator) 
along with selection of parameters of single-particle and pair fields is given 
in Refs~\cite{LM95,LM01}, so the present consideration is confined to description 
of its features as applied to the OIF method.
Most equations are given in the framework of the approximations (\ref{I_x}) 
since the approximation is used in the calculations. 
The equations for the exact treatment of $I_a$ operators
in the semi-perturbative approximation are obtained in close analogy with these ones.

Consider quasiparticle states with definite values of the angular momentum
projection $K$ (with definite values of $|K|$ and signature), 
which are obtained by a solution of HFB equations 
for the Hamiltonian $H$, and operators $b^\dagger_l$,
creating quasiparticles in these states. Within the simple model, $H$ contains only
a single-particle Hamiltonian, diagonal in the representation
of the particle operators $a^\dagger_l$, $a_l$, and pairing interaction, so
$b^\dagger_l =u_l a^\dagger_l +v_l a_{\bar{l}}$,
${\mathcal H}\,b^\dagger_l|0\rangle = \epsilon_l \,b^\dagger_l|0\rangle$, where
$\mathcal H$ is a quasiparticle Hamiltonian.
All the following formulas are not limited to the case of this simple Hamiltonian since,
with the corresponding replacement of $u_l,v_l$ by the coefficients of the general
Bogoliubov transformation, they are valid for $\mathcal H$,
obtained from any energy density functional.

In the semi-perturbative approximation, eigenfunctions of the Hamiltonian
$H_\omega $ (or, at next iterations, $\widetilde{H}_\omega$) are sought in the form
\begin{equation} \label{Psi}
\Psi = e^{iF_{\scriptstyle x} I_{\scriptstyle x}} \sum_l C_l b^\dagger_l|0\rangle  \;.
\end{equation}
For nucleus states with moderate spin $I$, the operator $F_x I_x$
may be treated as a small perturbation. It is useful to chose $F_x$
as one-particle operator, then the transformation 
$e^{iF_{\scriptstyle x} I_{\scriptstyle x}}$ 
does not change the expectation value $\langle N \rangle$. 
Expanding the transformed Routhian $H'_\omega$,
\begin{equation} \label{H'}
H'_\omega = e^{-iF_{\scriptstyle x} I_{\scriptstyle x}} H_\omega 
e^{iF_{\scriptstyle x} I_{\scriptstyle x}}\;,
\end{equation}
in powers of $iF_xI_x$ and neglecting terms of power higher than two,
we find operator $F_x$ from the equation $(H'_\omega)^{20} =0$
that involves only terms with $\Delta K=\pm 1$.
The terms with $\Delta K=0$, which do not depend on $F_x$, become 0 by the definition
of amplitudes $u,v$, while such terms, depending on $F_x$, are small.
Hereinafter, the denotations $A^{ij}$ are used for operators $A$ in representation
of operators $b^\dagger_l$. Amplitudes $C_l$ are found by the diagonalization of
the Hamiltonian $(H'_\omega)^{11}$ under the condition (\ref{J_x}) 
(see equations in Refs.~\cite{LM95,LM01}). 
As a result, the eigenvectors of this Hamiltonian have the form
$\beta^\dagger_p|0\rangle$, $\beta^\dagger_p \equiv \sum C_l(p)\, b^\dagger_l$.

Taking into account these approximations and the smallness of the ME
of the operator $F_x$, we get equations of the model. The ME of operator
$\theta^a_{lm}$ are found by the variation of the functional (\ref{F}),
in which the expectation values are calculated over the state
$|s\rangle \equiv \beta^\dagger_s|0\rangle$, corresponding to the nuclear level
in question. Since in the leading term of the functional,
i.~e. in $ \langle H \rangle$, we take into consideration only the single-quasiparticle
part of the Hamiltonian $H$, the same approximation is made also in terms
of the next order of smallness, i.~e., in operators, containing
$\theta$ and $F_x$. Then, the variation of the functional (\ref{F}) leads to the set
of linear equations for $\langle l|\theta_x|m\rangle$ (with $l<m$)
and $\eta_x$:
\begin{eqnarray}
&& 2 \left\{ [ 2 - C^2_l - C^2_m ] (\epsilon_l + \epsilon_m) \left( L^+_{lm}
\right) ^2 \right. \left. \,+\, \left( C^2_l - C^2_m \right) (\epsilon_m -
\epsilon_l) \left( M^-_{lm} \right) ^2 \right\} \, i\theta^x_{lm} \label{theta^x} \\
&& - \sum_{k(k \neq l,m)} \, C_k \, C_l \, \left\{ \, L^+_{lm} \, L^+_{km} \,
(2\epsilon_m + \epsilon_k + \epsilon_l) \right. \left. \, + \, M^-_{lm}
M^-_{km} (\epsilon_k + \epsilon_l - 2\epsilon_m) \, \right\} \,
i\theta^x_{km}  \notag \\
&& \, - \, \sum_{k(k \neq l,m)} \, C_k \, C_m \, \left\{ \, L^+_{lm} \,
L^+_{lk} \, (2\epsilon_l + \epsilon_k + \epsilon_m) \right. \left. \, + \,
M^-_{lm} M^-_{lk} (\epsilon_m + \epsilon_k - 2\epsilon_l) \, \right\} \,
i\theta^x_{lk}  \notag \\
&& \, - \, \frac{1}{\langle (\Delta J_x)^2\rangle} \, h_{lm} \eta_x \, =
\, B_{lm} \,. \notag
\end{eqnarray}
Here and below the following denotations are used for brevity:
$\theta^x_{lm} = \langle l|\theta_x|m\rangle$, 
$j^x_{lm} = \langle l|j_x|m\rangle$,
$M^\pm_{lm} = (u_lu_m \mp v_lv_m)$, 
$L^\pm_{lm} = (u_lv_m \pm v_lu_m)$,
$J^{20}_{l\bar{m}} = j^x_{lm} L^-_{lm}$,
$J^{11}_{lm} = j^x_{lm}M^+_{lm}$,
$\theta^{20}_{l\bar{m}} = \theta^x_{lm} L^+_{lm}$, $C_l = C_l(s)$,
\begin{eqnarray}
h_{lm} &=& 2 \left\{\, 2 J^{20}_{l\bar{m}} \, L^+_{lm} \,-\, C_l \,
\sum_k \, C_k \, M^-_{kl} \, j^x_{km} \right. \left. \,+\, C_m\, \sum_k \,
C_k \, M^-_{km} \, j^x_{kl} \, \right\} \,, \label{h_lm} \\
B_{lm}& = & \frac{1}{\langle (\Delta J_x)^2\rangle}\, \left\{ \, L^+_{lm}
\,(\epsilon_l + \epsilon_m) \, \left[ \, 2 \, J^{20}_{l\bar{m}} \,-\, C_l \,
\sum_k \, C_k \, J^{20}_{k\bar{m}}
-\, C_m \,\sum_k \, C_k \, J^{20}_{l\bar{k}} \, \right] \right. \label{B_lm}\\
&-& \left. {}M^-_{lm}\, (\epsilon_m + \epsilon_l)\,
\left[ \, C_l \, \sum_k \, C_k \, J^{11}_{km}
\,+\, C_m \, \sum_k \, C_k \, J^{11}_{lk}
\,-\, 2 C_l\, C_m \, \langle s|J_x|s\rangle \, \right]\, \right\} \,. \notag
\end{eqnarray}
The set is appended with equation (\ref{<J,theta>}),
which in the said approximation can be written as
\begin{eqnarray}
&& \sum_{0<l<m} \, h_{lm}\, i\theta^x_{lm} \,+\,
\frac{{\mathcal J}^{\text{unmix}}}{\langle (\Delta J_x)^2\rangle} \, \eta_x
\,=\, 1 \,-\, \frac{Z^{\text{unmix}}}{\langle (\Delta J_x)^2\rangle} \,, \label{sum}\\
&& Z^{\text{unmix}} \equiv 2 \sum_{0<l<m} \, \left| \, J^{20}_{l\bar{m}} \, \right| ^2
\,,
\quad {\mathcal J}^{\text{unmix}} \equiv 4\sum_{0<l<m}
\frac{\left| \, J^{20}_{l\bar{m}} \, \right| ^2}{(\epsilon_l + \epsilon_m)}
\,. \label{unmix}
\end{eqnarray}
Summation in quantities $Z^{\text{unmix}}$ and ${\mathcal J}^{\text{unmix}}$
is carried out only over those single-particle states
(with positive signature) that do not take part in the Coriolis mixing 
of states, occupied by an odd nucleon.
The "unmixed" are the states that have parity or isospin projection,
alternative to the corresponding properties of the investigated state $|s\rangle $.
For them, $C_l(s) = C_m(s)=0$ and Eq.~ (\ref{theta^x}) take a simple form
\begin{equation}
i\theta^{20}_{l\bar{m}} = \frac{1}{\langle (\Delta J_x)^2\rangle}
\left( 1+ \frac{2 \eta_x}{\epsilon_l + \epsilon_m} \right) J^{20}_{l\bar{m}}\, ,
\label{theta_20_lm}
\end{equation}
which was used in the derivation of Eq.~(\ref{sum}). One can included into
the set of "unmixed" states also non-alternative states, which are not involved
in the basis for calculation of the amplitudes $C_l(s)$.

Similar equations are obtained for $\theta^y_{lm}$ and $\eta_y$.
After solving of all these equation, the eigenfunction of the Hamiltonian 
$\widetilde{H}_\omega$ are sought in the form (\ref{Psi}). The matrix elements
of the operator $F_x$ are determined by the equation
\begin{equation}
\left[H,iF^{20}_x \right]I_x -\omega_{\text{tot}} J^{20}_x = 0\, , \quad
\omega_{\text{tot}} \equiv \omega + \langle s|[ H,i\theta_x ]|s \rangle 
+  \langle s|[[ H,i\theta_x ],iF_x ]|s \rangle \, I_x
\label{F_20_lm}
\end{equation}
with the obvious solution 
$i(F_x)^{20}_{l\bar{m}} I_x = \omega_{\text{tot}} J^{20}_{l\bar{m}} /
(\epsilon_l + \epsilon_m) $.
The amplitudes $C_l(s)$ are found by the diagonalization of the Hamiltonian
\begin{eqnarray} \label{tildeH_11}
\left(
e^{-iF_{\scriptstyle x} I_{\scriptstyle x}} \widetilde{H}_\omega 
e^{iF_{\scriptstyle x} I_{\scriptstyle x}} \right)^{11}
&=&
\widetilde{H}^{11}
- \Bigl(
\omega_{\text{tot}} - \langle s|[H,i\theta_x]|s \rangle 
+  \langle s|[[H,i\theta_x ],i\theta_x ]|s \rangle \, \langle s|J_x|s \rangle 
\Bigr) J^{11}_x \\
&+& [H,i\theta^{11}_x] \, \langle s|J_x|s \rangle 
+ [H,iF^{11}_x ] \, I_x - \frac{1}{2}\,\omega_{\text{tot}}^2 {\mathcal J}^{11}
- \frac{\omega_{\text{tot}} J^2_z} {2\sqrt{I(I+1)-\langle J^2_z\rangle}}\,,
\notag \\
(F_x)^{11}_{lm} &=& \frac{(F_x)^{20}_{l\bar{m}}}{L^+_{lm}} M^-_{lm}, \quad
{\mathcal J}^{0+11} \equiv \sum_{lm}
\frac{\left|\, J^{20}_{l\bar{m}}\, \right| ^2}{(\epsilon_l + \epsilon_m)}
- 2 \sum_{lm}
\frac{\left|\, J^{20}_{l\bar{m}}\, \right| ^2}{(\epsilon_l + \epsilon_m)}
b^\dagger_l b_l
\end{eqnarray}
The amplitudes are determined self-consistently for each value of 
$\omega_{\text{tot}}$ since the Hamiltonian itself 
depends on $|s \rangle $. One can easy check that, after 
the $e^{iF_{\scriptstyle x} I_{\scriptstyle x}}$ transformation,
the Eq.~(\ref{J_x}) depends on $\omega_{\text{tot}}$ (and on the amplitudes $C_l(s)$).
Therefore it is comfortable, instead of $\omega$, to find the solution 
$\omega_{\text{tot}}$ to this equation. When $\omega_{\text{tot}}$
and $C_l(s)$ have been obtained, all the procedure for $\theta$ and $C_l(s)$
is repeated until the convergence of the iterative process is attained.
The energy is calculated  as $\langle \widetilde{H} \rangle$ (see Eq.~\ref{F}).
One can note that, in the given model, the Hamiltonian (\ref{tildeH_11}) 
does not contain ME $\theta^y_{lm}$, while $\langle \widetilde{H} \rangle$ 
depends on these values.
\begin{figure}[ptb] 
\begin{center}
\includegraphics[scale=0.65]{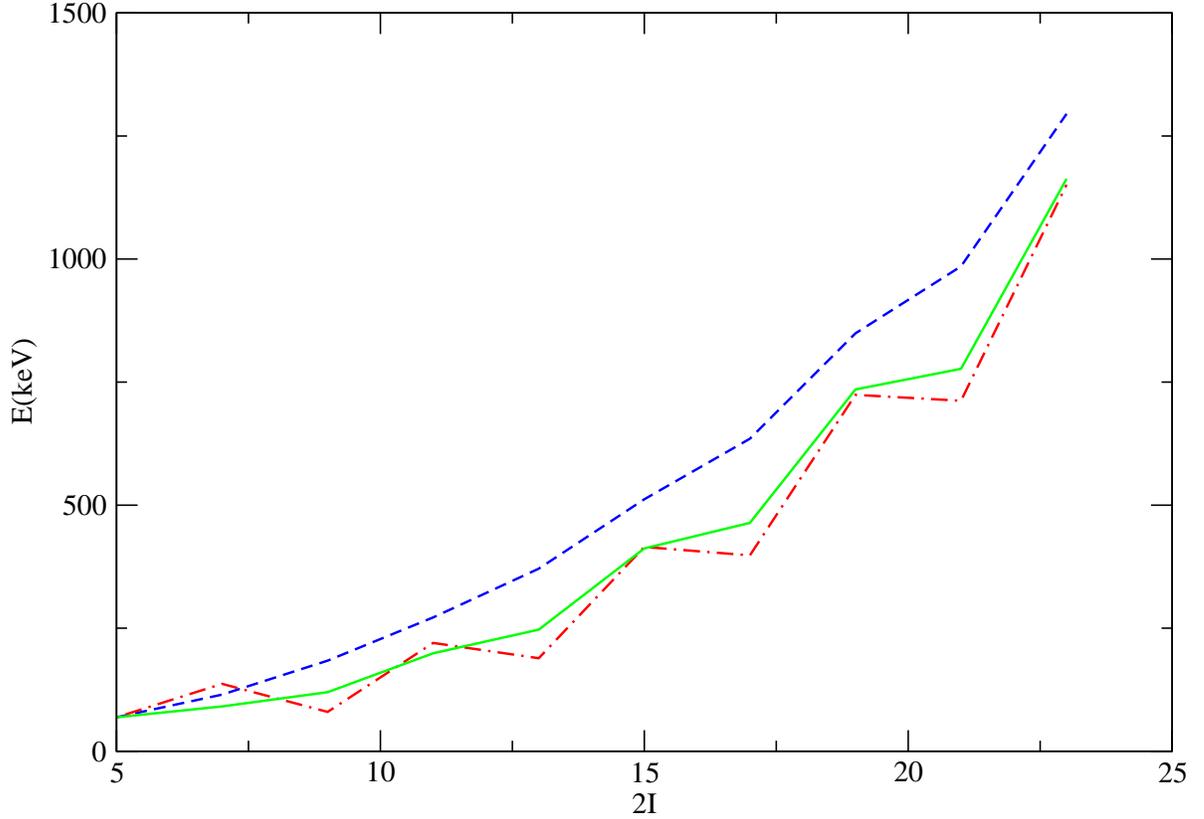}
\end{center}
\caption{\label{fig1} Energies of levels of the rotational band 
5/2~$^+$[642] in $^{163}$Er.
The dashed blue and the dash-dot red lines correspond to the results of calculations
in ordinary CM and in the semi-perturbative version of the OIF method.
The solid green line is the experimental data (\cite{NDS}).}%
\end{figure}
Let us make notes about application of the semi-perturbative approximation
in the full-scale self-consistent calculations with the exact treatment 
of the operators $I_a$. As it was mentioned in Sect.~\ref{sect2}, 
this treatment is usefull in calculations for band-head levels 
and closed to them ones.
In this case, the quasiparticle states $b^\dagger_l|0\rangle$
with definite values of $|K|$ and signature are found for the
Hamiltonian ${\mathcal H}$ depending on the operators $\theta_a$.
The HFB equation for this Hamiltonian are obtained by the minimization
of the functional (\ref{F}) taken without of the terms containig 
the operators $I_a$. Then the transformation (\ref{Psi}) is used,
where the operator $F_xI_x$ is changed by $\sum F_aI_a$ and
the operators $F_a$ are determined by elimination of the terms
$ \widetilde{H}^{20}_{\mu}$. The Lagrange multiplier $\mu$
in the Hamiltonian $ \widetilde{H}_{\mu}= \widetilde{H}-\mu \sum I_aJ_a$
is defined by Eq.~(\ref{<IJ>}).

Return to the simple model. 
The Hamiltonian is selected as the Nilsson potential with constant
pairing. In this calculations, we will consider only rotational
level energies. At present, the theory (particularly, in such
simple form) predicts intrinsic motion energies with poor
accuracy, which can tell on the description of rotational bands
both for the worse and for the better. To render the
description of rotational states independent of the quality of
calculations of intrinsic motion energies, these energies are
found by fitting of band-head energies to experiment, with
theoretical values taken for those unavailable from experiment.
It should be pointed out, however, that all rotational bands of
odd nuclei (except for band-head states) are calculated without
fitted parameters.

Results of calculations of energies of levels for the rotational band
with strong mixing in $^{163}$Er are presented in Fig.~\ref{fig1}.
Here, dashed blue, dash-dot red and solid green lines
denote, respectively, energies obtained without and with account
of rotational recoil operator, and experimental values (\cite{NDS}).
As seen from the Figure, taking into account of the effect even
in the relatively simple version of the proposed model noticeably improves
the agreement of theory with experiment for the rotational band with strong mixing.
The mixing for negative parity bands of $^{163}$Er is less pronounced
and so the role of the effect for energies, counted off from band-head levels,
is less than for positive parity ones.
For all bands, however, the rotational recoil term contributes much to the
energy of intrinsic motion; therefore, to take correct account of it
is of great importance in self-consistent calculations of all quantities
in deformed nuclei.
\section{Conclusions \label{sect5}}
%
The present work is an attempt to remove some difficulties arising 
in the microscopic description of deformed nuclei which were mentioned
in the Introduction and to develop the method of the intrinsic frame
which is considered today as too complex and not elaborated for calculations. 
For this purpose, the method of the OIF has been developed
on the base of the Mikhajlov concept of the unitary transformation to the IF.
The method seems to be more complex in its justification and development
but may be simpler in applications than the AMP one, for example, 
in the self-consistent description of the quasiparticle-rotation coupling.
Though the obtained results are intended for odd nuclei, most part of them can 
also be useful for even ones.

Some of deficiencies of the work and ways of feather
development are obvious. First of all, it is desirable to perform microscopic
calculations with effective forces based on contemporary energy density 
functionals. One can hope that such calculations may be applied 
not only for the description of deformed nuclei but also 
for the study of the effective interactions,
exploiting for this purpose quasiparticle excitations, 
and the fact that the single-particle motion is much less 
disturbed by particle in these nuclei than in spherical ones. 
For the second, the method should be implemented and further developed 
for calculations of transition probabilities.
Next, small deviations from axial symmetry can be considered in the frame of the given
version of  OIF by the mixing of states, nevertheless, it is desirable to study high
asymmetries using the 3D transformation and the corresponding constrains
in the IF. Also, it should be mentioned that there are many vibrational and
quasiparticle-vibrational states in the deformed nuclei 
while the present concept of the OIF is based on the HFB
method (or on the semi-perturbative approximation), 
so the development of the method for a self-consistent description 
of these states is the problem of future.

\begin{acknowledgments}
Helpful discussions with V. M. Mikhajlov are gratefully acknowledged.
This work has been supported by the Deutsche Forschungsgemeinschaft under
the grant No.~436 RUS113/994/0-1 and by the Russian Foundation for Basic
Research under the grant No.~09-02-91352-DFG\_a.
\end{acknowledgments} 


\begin{thebibliography}{99}
\bibitem{BMII} A. Bohr and B. R. Mottelson, {\it Nuclear Structure},
Vol. II, (W. A. Benjamin Inc., New York, 1975).
%
\bibitem{EG} J. M. Eisenberg and W. Greiner, {\it Nuclear Theory}, Vol. 3: 
{\it Microscopic Theoyr of the Nucleus} (North Holland, Amsterdam, 1973).
%
\bibitem{RS} P. Ring and P. Schuck, {\it The Nuclear Many-Body Problem}
(Springer-Verlag, Heidelberg, 1980).
\bibitem{Ring_1985} P. Ring in {\it Winter College on Fundamental Nuclear Physics.
Trieste}, 1984, Vol. 2, edited by K. Dietrich, M. Di Toro and H. J. Mang
(World Scientific, Singapore, 1985) p. 799.
\bibitem{Schmid_Gruemmer} K. W. Schmid and F. Gr\"{u}mmer,
Rep. Prog. Phys. {\bf 50}, 731 (1987).
\bibitem{Hara_1995} K. Hara and Y. Sun, Int. J. Mod. Phys. {\bf E4}, 637 (1995).
\bibitem{Frauendorf_2001} S. Frauendorf, Rev. Mod. Phys. {\bf 73}, 121 (2001).
\bibitem{Bender_RMP_2003} M. Bender, P. H. Heenen, and P.-G. Reinhard,
Rev. Mod. Phys. {\bf 75}, 121 (2003).
\bibitem{Egido_LNP_2004} J. L. Egido and L. M. Robledo,
Lect. Notes Phys. (Springer, Berlin/Heidelberg) {\bf 641}, 269 (2004).
\bibitem{Schmid_2004} K. W. Schmid, Progr. Part. Nucl. Phys. {\bf 52}, 565 (2004).
\bibitem{Caurier_RMV_2005} E. Caurier, G. Martinez-Pinedo, F. Nowacki,
A. Poves, and A. P. Zuker, Rev. Mod. Phys. {\bf 77}, 427 (2005).
\bibitem{Satula_Wyss_2005} W. Satu{\l}a and R. A. Wyss, Rep. Prog. Phys. 
{\bf 68} (2005) 131.
\bibitem{Vretenar_2005} D. Vretenar, A. V. Afanasjev, G. A. Lalazissis, and P. Ring,
Phys. Rep. {\bf 409}, 101 (2005).
\bibitem{Zalewski_2008} M. Zalewski, J. Dobaczewski, W. Satu{\l}a, 
and T. R. Werner, Phys. Rev. {\bf C 77}, 024316 (2008).
\bibitem{Fujii_2009} S. Fujii, R. Okamoto, and K. Suzuki, 
Phys. Rev. Lett. {\bf103}, 182501 (2009).
\bibitem{Bortignon_2010} P. F. Bortignon, G. Colo and H. Sagawa, 
J. Phys. G: Nucl. Part. Phys. {\bf 37}, 064013 (2010).
\bibitem{Zalewski_2010} M. Zalewski, P. Olbratowski, and W. Satu{\l}a,
Phys. Rev. {\bf C 81}, 044314 (2010).
\bibitem{Barbieri_2009}
S. J. Waldecker, C. Barbieri, and W. H. Dickhoff, 
arXiv:1105.4257v1 [nucl-th];
C. Barbieri, arXiv:0909.0336v2 [nucl-th]
\bibitem{Niksic_2008} T. Nik\v si\'c, D. Vretenar, and P. Ring. Phys. Rev. {\bf C 78},
034318 (2008).
\bibitem{Baldo_2010} M. Baldo, L. M. Robledo, P. Schuck, and X. Vinas,
arXiv:1005.1810v1 [nucl-th].
\bibitem{Hinohara_2008} N. Hinohara, T. Nakatsukasa, M. Matsuo, and K. Matsuyanagi,
Prog.Theor.Phys. {\bf 119}, 59 (2008).
\bibitem{Shimizu_2008} Y. R. Shimizu and K. Matsuyanagi, arXiv:nucl-th/0008005v1.
%
\bibitem{Nomura_2011}Kosuke Nomura, Takaharu Otsuka, Noritaka Shimizu, and Lu Guo,
Phys. Rev. {\bf C 83}, 041302(R) (2011)
\bibitem{Vargas_2002} C. E. Vargas, J. G. Hirsch, and J. P. Draayer, 
Phys. Rev. {\bf C 66}, 064309 (2002).
%
%
\bibitem{Bender_2008} M. Bender and P.-H. Heenen, Phys. Rev. {\bf C 78}, 024309 (2008)
%
\bibitem{Delaroche_2010} J.P. Delaroche, M. Girod, J. Libert, H. Goutte, 
S. Hilaire, S. Peru, N. Pillet, and G.F. Bertsch, Phys. Rev. {\bf C 81}, 014303 (2010)
\bibitem{Girod_2009} M. Girod, J.-P. Delaroche, A. Goergen, and A. Obertelli,
Phys. Lett. {\bf B 676}, 39 (2009).
\bibitem{Rodriguez_2011} Tomas R. Rodrguez and J. Luis Egido,
arXiv:1109.6516v2 [nucl-th].
\bibitem{Yao_2009} J. M. Yao, J. Meng, P. Ring, and D. Pena Arteaga,
Phys. Rev. {\bf C 79}, 044312 (2009).
\bibitem{Ring_2011} P. Ring, H. Abusara, A. V. Afanasjev, G. A. Lalazissis, 
T. Nik\v si\'c, and D. Vretenar
Int. J. Mod. Phys. {\bf E 20}, 253 (2011).
%
\bibitem{Bes86} D. R. Bes, H. T. Mehr, and P. J.Liotta, Nucl. Phys.
{\bf A449}, 459 (1986).
\bibitem{Bes91} D. R. Bes, R. De Luca, Nucl. Phys. {\bf A535}, 221 (1991).
%
\bibitem{LM98} N. A. Lyutorovich and V. M. Mikhajlov, Izv. Ross. Akad. Nauk, 
Ser. Fiz. {\bf 62}, 923 (1998) [Bull. Russ. Acad. Sci., Phys. {\bf 62}, 744 (1998)].
%
\bibitem{Peng_2008} J. Peng, J. Meng, P. Ring, and S. Q. Zhang,
Phys. Rev. {\bf C 78}, 024313 (2008)
\bibitem{Shoji_2009} T. Shoji, Y. R. Shimizu, Prog. Theor. Phys. {\bf 121}, 319 (2009).
%
\bibitem{Dobaczewski_2009} J. Dobaczewski, W. Satu{\l}a, B. G. Carlsson, J. Engel, 
P. Olbratowski, P. Powalowski, M. Sadziak, J. Sarich, N. Schunck, A. Staszczak, 
M. Stoitsov, M. Zalewski, and H. Zdunczuk, 
Comput. Phys. Commun., {\bf 180}, (2009) 2361.
%
\bibitem{Schunck_2011} N. Schunck, J. Dobaczewski, J. McDonnell, W. Satula, 
J.A. Sheikh, A. Staszczak, M. Stoitsov, P. Toivanen, arXiv:1103.1851v2 [nucl-th], 
Comput. Phys. Commun., In Press, Available online 27 August 2011.
%
\bibitem{Carlsson_2007} B. G. Carlsson, Inter. Journ. Mod. Phys. {\bf E 16} (2007) 634.
\bibitem{Qi_2009} B. Qi, S. Q. Zhang, J. Meng, S. Y. Wang, and S. Frauendorf
Phys. Lett. {\bf B 675}, (2009) 175. 
\bibitem{Kamlah}A. Kamlah, Z. Physik {\bf 216}, 52 (1968).
\bibitem{Hilton} R. R. Hilton, H. J. Mang, P. Ring, J. L. Egido, H. Herold, M. Reinecke, 
H. Ruder, and G. Wunner, Nucl. Phys. {\bf A 366}, 365 (1981).
\bibitem{Frauendorf_1969} S. Frauendorf, J. Janssen, and I. Munchov,
Nucl. Phys. {\bf A125}, 369 (1969).
\bibitem{Jansen_1979} D. Jansen and I. N. Mikhailov, Nucl. Phys. {\bf A318}, 390 (1979).
\bibitem{Une_2001} T. Une, Prog. Theor. Phys. {\bf 106}, 941 (2001).
\bibitem{Zdunczuk_2007} H. Zdu\'nczuk, W. Satula, J. Dobaczewski, and M. Kosmulski
Phys. Rev. {\bf C 76}, 044304 (2007).
\bibitem{LM95} N. A. Lyutorovich and V. M. Mikhajlov, Phys. Lett. 
{\bf B 356}, 163 (1995).
\bibitem{Ring_1974} P. Ring, H. J. Mang, and B. Banerjee, Nucl. Phys. {\bf A225}, 
141 (1974)
\bibitem{Gareev}F. A. Gareev, S. P. Ivanova, V. G. Soloviev, and S.I. Fedotov.
Phys. Element. Part. Atom. Nucl. {\bf 4}, 357 (1973).
\bibitem{Kvasil} L. Kvasil, I. N. Mikhailov, R. Ch. Safarov, and B. Choriev,
Czech. J. Phys. {\bf B28}, 843 (1978).
%
\bibitem{Mikhailov_71} V. M. Mikhajlov, Izv. Akad. Nauk SSSR, 
Ser. Fiz. {\bf 35}, 794 (1971) [Bull. Acad. Sci. USSR, Phys. {\bf 35}, 794 (1971)].
%
\bibitem{Mikh93} V. M. Mikhajlov, in {it Thesises of 44-th Conference on Nuclear
Spectroscopy and Atomic Nucleus Structure} (St. Petersburg, 1994) p. 132.
%
\bibitem{ML94} V.M. Mikhailov and N.A. Lyutorovich, Izv. Ross. Akad. Nauk, 
Ser. Fiz. {\bf 58}, No. 11, 24 (1994)
[Bull. Russ. Acad. Sci., Phys. {\bf 58}, No. 11, 24 (1994)].
%
\bibitem{Varshalovich} D. A. Varshalovich, A. N. Moskalev, and V. K. Khersonskii,
{\it Quantum Theory of Angular Momentum} (World Scientific, Singapore, 1988).
%
\bibitem{LM91}  N. A. Lyutorovich and V. M. Mikhailov, Izv. Ross. Akad. Nauk, 
Ser. Fiz. {\bf 55}, 2214 (1991) [Bull. Russ. Acad. Sci., Phys. {\bf 55}, 2214 (1991)].
%
\bibitem{Dobaczewski_1996} J. Dobaczewski, W. Nazarewicz, T. R. Werner, 
J.-F. Berger, C. R. Chinn, and J. Decharg\'e, Phys. Rev. {\bf C 53}, 2809 (1996).
%
\bibitem{LM01} N. A. Lyutorovich and V. M. Mikhailov, Izv. Ross. Akad. Nauk, 
Ser. Fiz. {\bf 65}, 634 (2001) [Bull. Russ. Acad. Sci., Phys. {\bf 65}, 681 (2001)].
%
\bibitem{NDS} C. W. Reich, Nucl. Data Sheets {\bf 111}, 1211 (2010) 
%
\end{thebibliography}

\end{document}